\documentclass{article}

\usepackage{amsmath}
\usepackage{graphicx}
\usepackage{psfrag}
\def\be{\begin{equation}}
\def\ee{\end{equation}}
\def\bea{\begin{eqnarray}}
\def\eea{\end{eqnarray}}
\def\bml{\begin{mathletters}}
\def\eml{\end{mathletters}}

\begin{document}
%=============================================================================
%=============================================================================
\title{Nonlinear deterministic equations in biological evolution}

\author{Kavita Jain\footnote{also at Evolutionary and Organismal Biology Unit} and Sarada Seetharaman \\  
Theoretical Sciences Unit \\ 
Jawaharlal Nehru Centre for Advanced Scientific Research, \\
Jakkur P.O., Bangalore, India \\
{\tt jain@jncasr.ac.in, saradas@jncasr.ac.in}}

\label{firstpage}
\date{\today}

\maketitle

\begin{abstract}
We review models of biological evolution in which the population
frequency changes deterministically with time. If the population is
self-replicating, although the equations for simple prototypes can be
linearised,  nonlinear equations arise in many complex situations. For
sexual populations, even in the simplest setting, the 
equations are necessarily 
nonlinear due to the mixing of the parental genetic material. The
solutions of such nonlinear equations display interesting features such as multiple equilibria and phase
transitions. We
mainly discuss those models for which an analytical 
understanding of such nonlinear equations is available.
\end{abstract}

%=============================================================================
%INTRODUCTION
%=============================================================================
\section{Introduction}

A population evolves when the changes that happen  during a generation are
passed on to the subsequent generations. These changes may happen in
the somatic immune cells in order to adapt to a microbe attack or in
the germline cells. Though in both the cases the genome is altered, in
the former, it also manifests as changes in the composition of the
protein coded by that part of the genome. Therefore one defines the 
models describing biological evolution in genotype or
protein space \cite{Smith:1970}.  

The quantity of interest is the population frequency of a genotype
which changes under the action of two elementary processes namely
selection and mutation. In the simplest setting, the time-dependent
equations for the population fraction are nonlinear but they can be
linearised and the steady state solution obtained at long times
 can be shown to be unique. In more complex situations such as when
 subpopulations are 
coupled to each other or when the growth rate of a
genotype depends on its current frequency, nonlinear evolution
equations give rise to  multiple equilibria. In the cases where the 
solution is unique, phase transition may occur in the
steady state.   
If the process of sexual reproduction is also included, the resulting
equations are 
bilinear in population and such inherently nonlinear equations exhibit
multiple solutions in the steady state and dynamic phase transitions.

In this review, we will focus on infinite populations which obey
deterministic equations of evolution. Although the real populations
are finite and evolve stochastically,  
phenomena observed in deterministic setting may survive in the presence of
stochasticity as 
well \cite{Quer:1996}, and deterministic solutions can also be utilised to get
insight in the corresponding stochastic problem \cite{Jain:2007a} and 
to develop stochastic theories \cite{Jain:2008b}. 
For a discussion of topics not covered in this article, we refer the
reader to several excellent textbooks
\cite{Ewens:1979,Durrett:2002,Nowak:2006} and other review articles on
the subject \cite{Jain:2007b}.  

The article is organised as follows. In the next section, we introduce some 
 basic concepts and definitions. This is followed by a discussion of
 models for asexually reproducing populations in Sec.~\ref{asex} and
 sexually reproducing ones in Sec.~\ref{sex}. Finally a 
summary and outlook is presented in Sec.~\ref{sum}.
%=============================================================================
%SETUP
%=============================================================================
\section{Basic definitions}
\label{basic}

In this section, we explain some basic concepts and definitions which
are relevant to the discussion in the following sections.

\begin{figure} 
\begin{center}
\begin{small}\begin{footnotesize}             \end{footnotesize}\end{small}\includegraphics[width=0.4\linewidth,angle=0]{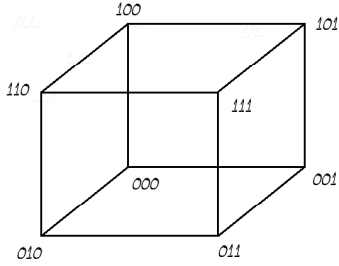} 
\caption{The sequence space for $L=3$ represented on a
  Hamming cube. }
\label{hamming}
\end{center}
\end{figure}

{\it Sequence and sequence space:} A sequence
$\sigma=\{\sigma_1,...,\sigma_L \}$ is a string of $L$ letters which
are chosen from an alphabet of size $a$. 
It represents a protein if $\sigma_i$ denotes one of the
$a=20$ amino acids and a genotype when the
letters are one of the four nucleotides. 
The total sequence space consists of all possible
strings of length $L$  and thus has
a size $n=a^L$  which increases exponentially with $L$. For
  computational ease, it is useful to lump some of the information in
  a single letter. For example, instead of working with all the four
  nucleotides in a genotype, one can classify them as 
  purines (adenine and guanine) and pyrimidines (thymine and cytosine)
  thus reducing $a$ to two. Similarly instead of considering all
  possible mutations at a locus, one may differentiate between
  genotypes by the absence or presence of a
  mutation which again corresponds to $a=2$ \cite{Visser:2009}. In
  this article, we will work with binary 
  sequences unless specified otherwise. Such 
  $n=2^L$ sequences can be arranged on a Hamming space, 
an example of which is shown in Fig.~\ref{hamming} for binary sequence
of length $L=3$. Two sequences $\sigma$ and $\sigma'$ are said to be at
Hamming distance $d(\sigma,\sigma')$ if they differ at $d$ loci. For a
binary sequence in which $\sigma_i=0$ or $1$, one may write
\be
d(\sigma,\sigma') = \sum_{i=1}^L (\sigma_i-\sigma'_i)^2
\ee

{\it Fitness:} The fitness $W(\sigma)$ of a sequence $\sigma$ is a 
measure of its reproductive success in a given environment. It 
represents the replication rate of a genotype or the functionality of a 
protein. The sequence space along with the 
fitness of each sequence comprises the {\it fitness
landscape}.  The choice of fitness landscape plays an important role in
determining the course of evolution and can be 
made according to the biological situation that one wishes to model
and the available experimental data or the analytical tractability of
the problem. 
A fitness landscape can be simple in that the fitness 
of a sequence depends only on its  distance  from a given
sequence. More often however the fitness landscapes are complex and
one has to specify all the $a^L$ fitnesses. These fitnesses can be
assumed to be independent random variables \cite{Jain:2005} or they
may have correlations \cite{Perelson:1995, Seetharaman:2010}. An
important feature of generic fitness landscapes is the presence of
epistasis which is a measure of the nonlinear contribution of  
locus fitness to the sequence \cite{Kouyos:2007}. If each locus
contributes independently to the sequence fitness, a  
 fitness landscape is said to be non-epistatic. Fitness can also
 depend on time as in
 the case of changing  environment \cite{Nilsson:2000,Nilsson:2002,Wilke:2001a} or
 it can be a 
 function of the concentration of the genotype frequency. In this
 review, we will employ various types of fitness landscapes.

{\it Mutation:} Stochastic changes known as mutations may
happen in the genome of 
an individual. These may insert, delete or change the
nucleotides in the genome and thus create a new sequence with a
different fitness. If the fitness of the mutant  is higher, the
change may propagate in the population and the population evolves 
towards a higher fitness value, otherwise it is eliminated.  In this
review, we will consider only {\it point mutations} that change a locus
$\sigma_i$ to one of the other $a-1$ possibilities with a certain
probability and thus preserve the 
length of the sequence.

{\it Recombination:} A sequence genetically different from the
parents can be produced by the recombination process in 
which two parent sequences mix to produce a 
new offspring sequence thus producing genetic variation within
  a population. Recombination occurs not only during gamete formation
  in sexually reproducing multicellular organisms but in unicellular organisms such as bacteria and fungi as well \cite{Hartl:2009}.  We will consider a recombination scheme 
(one-point crossover) in 
which the parent sequences $\sigma$ and $\sigma'$  break at a point
$i$ and exchange the genetic material with a certain probability
 resulting in offspring sequences $\{ \sigma_1,
...,\sigma_i, \sigma_{i+1}',...,\sigma_L' \}$ and $\{ \sigma_1',
...,\sigma_i', \sigma_{i+1},...,\sigma_L\}$.  

%=============================================================================
%ASEXUAL 
%=============================================================================
\section{Asexually reproducing populations}
\label{asex}

We first describe the equations governing the evolution of self
replicating populations.  Although the time-dependent equations for
the population frequency of such asexual populations are nonlinear in
general,  they can be linearised by a transformation of variables in
some simple 
cases \cite{Thompson:1974, Jones:1976}. We will
mainly discuss the steady state 
properties of these models in the following subsections. 

%--------------------------------------------------------------------------
\subsection{Haploid population}
\label{hap}

In a haploid population, each individual carries a single copy of its
genome sequence $\sigma$. In the presence of selection and mutation,
the  population frequency $X(\sigma,t+1)$ of a  sequence $\sigma$ at  
generation $t+1$ can be obtained from each sequence $\sigma'$ that makes
$W(\sigma')$ copies of itself in one generation and mutates to sequence
$\sigma$ with a probability $M(\sigma \leftarrow \sigma')$. This gives
the discrete time evolution equation as  
\be
X(\sigma,t+1)=\frac{\sum_{\sigma{'}} 
M(\sigma\leftarrow \sigma') W(\sigma{'})  X(\sigma{'},t)}
{\sum_{\sigma{'}} W(\sigma{'}) X(\sigma{'},t) } \label{exdis}
\ee
where the denominator on the right hand side (RHS) is the average fitness
${\cal W}(t)$ and ensures that the normalisation $\sum_{\sigma}
X(\sigma,t)=1$ is satisfied at all times.  If the mutation probability
per locus per 
generation is $\mu$ and  the point mutations occur
independently at each locus, the probability that a sequence $\sigma'$
mutates to sequence $\sigma$ at Hamming distance $d(\sigma,\sigma')$ is given
by
\be
M(\sigma \leftarrow \sigma')= \mu^{d(\sigma,\sigma')}
(1-\mu)^{L-d(\sigma,\sigma')}
\label{mutprob1}
\ee

It is evident that equation (\ref{exdis})  is nonlinear due to the
presence of denominator. However in terms of an unnormalised
population variable defined as 
\be
\label{Eigendisc}
Z(\sigma,t)=X(\sigma,t) \; \prod_{\tau=0}^{t-1} \sum_{\sigma{'}} W(\sigma{'}) 
X(\sigma{'},\tau)
\ee
we find that the unnormalised variables $Z(\sigma,t)$ obey a
linear equation given by  
\be
Z(\sigma,t+1)=\sum_{\sigma{'}} M(\sigma\leftarrow \sigma') W(\sigma{'}) Z(\sigma{'},t). \label{ezdis}
\ee
On writing
\be
X(\sigma,t)=\frac{Z(\sigma,t)}{\sum_{\sigma'} Z(\sigma',t)}
\label{transf}
\ee 
in (\ref{ezdis}), equation (\ref{exdis}) is obtained. 
In matrix notation, (\ref{ezdis}) can be written as ${\bf Z}(t+1)= A
{\bf Z}(t)$ where the $\sigma,\sigma'$ element of matrix A is given by
$M (\sigma \leftarrow \sigma') W(\sigma')$ and ${\bf Z}(t)$ is the
population vector at time $t$. Since the fitness $W(\sigma) \geq 0$,
the matrix $A$ is non-negative and it follows from the
Perron-Frobenius theorem that the largest eigenvalue of matrix $A$ is
real, positive and  nondegenerate with the corresponding eigenvector 
real and positive \cite{Bellman:1997}. Using this eigenvector in
(\ref{transf}) and taking the infinite time limit, the  
normalised frequencies in the steady state can be obtained. However in
some cases, it is possible to work directly with the nonlinear
equation (\ref{exdis}) in the steady state (see the discussion below).

In continuous time, one can write down the equation for the rate of
change $\dot{X}(\sigma,t) = \partial X(\sigma,t)/\partial t$ of the  
fraction $X(\sigma,t)$ of the population with sequence $\sigma$ as
\be
\dot{X}(\sigma,t)=\sum_{\sigma{'}} M(\sigma \leftarrow \sigma') 
W(\sigma{'}) 
X(\sigma{'},t)-\left(\sum_{\sigma{'}} W(\sigma{'}) X(\sigma{'},t) \right) 
X(\sigma,t) \label{exc}
\ee
where the last term on the RHS is the death term which accounts for
the normalisation $\sum_\sigma X(\sigma,t)=1$.  
Note that (\ref{exc}) is \textit{not}
the continuous time limit of (\ref{exdis}) although both equations
have the same steady state. 

The equations (\ref{exdis}) and (\ref{exc}) define respectively the discrete and continuous time versions of 
{\it Eigen's quasispecies model} \cite{Eigen:1971,Eigen:1977}. The main result
of the quasispecies theory is that  in the steady state, for several
choices of fitness landscapes, 
there exists a critical mutation rate below which the population forms
a quasispecies consisting of the fittest sequence and its closely
related mutants. Above this error threshold, the population is homogeneously
distributed over the entire sequence space. To illustrate this, we
consider the {\it sharp peak fitness landscape} defined by 
\be
W(\sigma)=W_0
\delta_{\sigma,{\bf 0}}+ (1-\delta_{\sigma,{\bf 0}})~,~W_0 > 1
\label{splh}
\ee
 where
${\bf 0}=\{0,0,...,0\}$ is the sequence with all zeros. Using this choice for $W(\sigma)$ in (\ref{exdis}) for the sequence ${\bf 0}$ in the steady state, we get
\be
X({\bf 0})= \frac{W_0 (1-\mu)^L X({\bf 0})+\sum_{\sigma{'} \neq {\bf 0}}
M(\bf 0\leftarrow \sigma') X(\sigma{'})}{W_0 X({\bf 0})+ 1-X({\bf 0})}
\ee
In the scaling limit $\mu \to 0, L \to \infty$ with $U=\mu L$ finite, the terms in the
numerator on RHS arising due to mutations to sequence ${\bf 0}$ vanish
and we obtain \cite{Nowak:1989} 
\be 
X({\bf 0}) = 1- \frac{U}{U_c} ~,~U < U_c=\ln W_0
\ee
Thus the master sequence ${\bf 0}$ supports a finite fraction of population below $U_c$. 
Above the critical probability $U_c$, the population is homogeneously distributed over the sequence space.

Not all fitness landscapes exhibit error threshold transition
\cite{Wiehe:1997}.  One such example is the 
non-epistatic {\it multiplicative fitness landscape} defined by
\be
W(\sigma)=\prod_{i=1}^L (1-s)^{\sigma_i}
\label{mult1}
\ee 
where the $0<s<1$ is a selection parameter. It can
be checked that the exact steady state frequency is  given by
\cite{Woodcock:1996},  
\be
X(\sigma)= \prod_{i=1}^L x_0^{1-\sigma_i} x_1^{\sigma_i}
\label{prod}
\ee
where $x_0, x_1$ are the
solutions of (\ref{exdis}) for the corresponding one locus model. For
a discussion of 
error threshold transition on other fitness landscapes, we refer 
the reader to \cite{Jain:2007b}.

If the replication and mutation  are treated as independent 
processes unlike in (\ref{exdis}) and
(\ref{exc}), we obtain the \textit{Crow-Kimura model} \cite{Crow:1970,
  Crow:1965} in which it is assumed that the replication process is
error-free and  
mutations occur due to external factors such as  radiation. Then the
equation for the 
rate of change  $\dot{X}(\sigma,t)$ 
can be written as \cite{Crow:1970, Akin:1979}
\be
\dot{X}(\sigma,t)=
[W(\sigma)-\sum_{\sigma'} W(\sigma') X(\sigma',t)] X(\sigma,t)+
\sum_{\sigma'} M(\sigma \leftarrow \sigma') X(\sigma',t). 
\label{ckxc}
\ee
where the mutation matrix is given by 
\bea
M(\sigma \leftarrow \sigma') = 
\begin{cases} 
0 & ~,~ \;\;d(\sigma,\sigma') > 1  \\
\mu & {~,~ \;\;d(\sigma,\sigma')=1 } \\
-L  \mu & {~,~ \;\;d(\sigma,\sigma')=0 }
\end{cases}
\label{mutprob2}
\eea
since $\sum_\sigma M(\sigma \leftarrow \sigma')$ should be zero.  
As in the Eigen's model, the nonlinearity in (\ref{ckxc}) can be eliminated by passing
to unnormalised population variables $Z(\sigma,t)$ defined by 
\be
Z(\sigma,t)= X(\sigma,t) \;\mbox{exp} \left[ \sum_{\sigma{'} }  W(\sigma{'}) 
\int_{0}^{t} d \tau X(\sigma{'},\tau) \right] \label{normzxc}
\ee
The error threshold transition for various fitness landscapes has been
demonstrated using the Crow-Kimura equation (\ref{ckxc}) also
\cite{Baake:2001, Saakian:2004a}. 

%--------------------------------------------------------------------------

\subsection{Diploid population}
\label{dip}

Higher organisms such as humans are diploid as they carry two copies
of their genome and we represent an individual of a diploid population
by $(\sigma,\sigma')$. A sequence is said to be homozygous if
  $\sigma$ and $\sigma'$ are identical and heterozygous otherwise. Selection-mutation equations analogous to the
haploid case can be written for  
the population frequency $X(\sigma,t)$ of the sequence $\sigma$. For the
Crow-Kimura model, the evolution 
equation reads as \cite{Wiehe:1995,Baake:1997a} 
\be
\dot{X}(\sigma,t)=
[{\tilde W}(\sigma,t)-\sum_{\sigma{''}} {\tilde W}(\sigma{''},t)
  X(\sigma{''},t)] X(\sigma,t)+ 
\sum_{\sigma'} M(\sigma \leftarrow \sigma') X(\sigma',t). 
\ee
where ${\tilde W}(\sigma,t)=\sum_{\sigma'}
W(\sigma,\sigma') X(\sigma',t)$ is the marginal fitness of sequence
$\sigma$ and $W(\sigma,\sigma')$ is the fitness of genotype
$(\sigma,\sigma')$.   
A transformation similar to (\ref{normzxc}) which can render the above
system of nonlinear equations linear is not known and the steady
solution may not be unique.  

The existence of multiple steady state solutions can be illustrated by
a diploid analogue of the sharp fitness landscape defined as  \cite{Wiehe:1995} 
\bea
\label{spldip1}
W({\bf 0},{\bf 0}) &=& f_0= 1+2 s \\
W({\bf 0},\sigma) &=& W(\sigma, {\bf 0})= f_1=1+2 h s ~,~ \sigma \neq
{\bf 0}\\
W(\sigma,\sigma') &=& 1 ~,~ \sigma, \sigma' \neq {\bf 0}
\label{spldip}
\eea
where $s, h > 0$. 
%\textbf{In the above equation, $s$ is the selectioncoefficient and
%$h$ is a dominance parameter that determines which of the two
%non-identical sequences would dominate the fitness of the
%heterozygote. When $h=0$, $W({\bf 0},\sigma)=W(\sigma,\sigma')=1$ and
%when $h=1$, $W({\bf 0},\sigma)=W({\bf 0},{\bf 0})=1+2 s$. In the case
%where $h=0.5,$ $W({\bf 0},\sigma)=1+s,$ the average of the wildtype's
%and mutant's fitness values.}
In the above equations, $s$ is a selection coefficient
  and
$h$ is a dominance parameter which controls the contribution of the master sequence to the fitness of the heterozygote. When $h=1$, since the fitness $W({\bf
    0},\sigma)=W({\bf 0},{\bf 0})$, the master sequence $\bf 0$ is dominant. On the other hand, when
  $h=0$, the fitness $W({\bf 0},\sigma)=W(\sigma,\sigma')=1$ and
  therefore the master sequence acts recessively. The dominance is
  absent when $h=1/2$ as the heterozygote fitness $W({\bf
  0},\sigma)=1+s$ is the average of the master fitness and the
  mutant fitness.
%When $h=0$, $W({\bf 0},\sigma)=W(\sigma,\sigma')=1$ and
%when $h=1$, $W({\bf 0},\sigma)=W({\bf 0},{\bf 0})=1+2 s$. In the case
%where $h=0.5,$ $W({\bf 0},\sigma)=1+s,$ the average of the wildtype's
%and mutant's fitness values.}

\begin{figure} 
\begin{center}
\includegraphics[width=0.5\linewidth,angle=270]{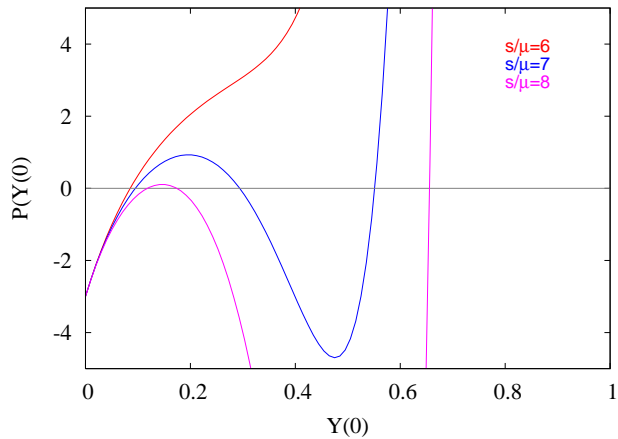} 
\caption{Plot of the polynomial $P(Y(0))$ as a function of $Y(0)$ for
  various $s/\mu$ (see
  Sec.~\ref{dip}). The equilibrium frequency $Y(0)$ is obtained when
  $P(Y(0))=0$.}
\label{diploid}
\end{center}
\end{figure}

 Using the above equation, the marginal fitness can
be written as  
\be
{\tilde W}(\sigma)= 
\begin{cases}
f_0 X({\bf 0})+f_1 (1-X({\bf 0})) ~,~\sigma={\bf 0} \\
f_1 X({\bf 0})+(1-X({\bf 0})) ~,~ \sigma \neq {\bf 0}
\end{cases}
\ee
and the average fitness as
\bea
\sum_{\sigma{''}} {\tilde W}(\sigma{''}) X(\sigma{''}) &= & X({\bf 0}) \left[f_0
X({\bf 0})+ f_1 (1-X({\bf 0})) \right] \nonumber \\
&+&  (1-X({\bf 0})) \left[f_1 X({\bf
    0})+ 1-X({\bf 0}) \right]
\eea
Since the fitness landscape (\ref{spldip1})-(\ref{spldip}) depends only on the Hamming distance from the
master sequence ${\bf 0 }$, one can work with the error
class frequencies $Y(d)$ which are obtained by summing over the
population fractions at Hamming distance $d$ from the master
sequence. 
Specialising to $h=0$, the steady state equation in terms of $Y$'s
reads as \cite{Wiehe:1995} 
\be
\sum_{d'=0}^L {\tilde M} (d \leftarrow d') Y(d') = 2 s Y^2(0) 
\begin{cases}
Y(0) -1 ~,~ d=0 \\
Y(d) ~,~ d \neq 0
\end{cases}
\ee
where the mutation matrix ${\tilde M}$ can be found using
(\ref{mutprob2}). The frequency $Y(0)$ obeys a polynomial
  equation of degree at most $2 (L+1)$. For small $L$, the above set
  of nonlinear equations can be straightforwardly 
solved. For $L=4$, the fraction $Y(0)$ obeys a
polynomial equation $P(Y(0))=0$ of degree $9$ \cite{Wiehe:1995}. The polynomial
$P(Y(0))$ is plotted against $Y(0)$ in Fig.~\ref{diploid} for various
$s/\mu$  to show the occurrence of multiple steady state
solutions. Which of these multiple solutions occur depends on 
  the initial conditions. For example, an initial distribution with
  $Y(0)=1$ gives 
  different steady state fitness from the initial condition $Y(L)=1$ 
   \cite{Wiehe:1995}.

%--------------------------------------------------------------------------

\subsection{Concentration-dependent fitness}
\label{concdep}

The fitness of a sequence is not always a constant and may depend
on the concentration of other sequences. In such cases, one ends
up with nonlinear dynamical equations which cannot be linearised. An
example of  
this scenario is the evolution of grammar in a population
\cite{Komarova:2001}. It 
has been proposed \cite{Sorace:1999,Nowak:2001} that  a set of grammars
${G_{1},G_{2},...,G_{n}}$ are innately available to  a learner and 
the language is learnt by just listening to the sentences and choosing the
correct grammar. 

A grammar that is easily understandable has a greater probability of
being propagated than the others and hence the fitness 
indicates its prevalence in the population. This is equal to the
fraction of sentences and their corresponding meanings that is common
between that grammar 
and all others multiplied by the population fraction using each
grammar.  
If $w(i,j)$ is the probability that a speaker of grammar $G_j$ can
understand a sentence by a user of grammar $G_i$, the fitness
$W(\{X(i)\})$  of grammar $G_i$ can be given as \cite{Komarova:2001}
\be
W(\{X(i)\})=\frac{1}{2} \sum_{j=1}^n\left[ w(i,j)+w(j,i) \right] X(j)
\ee
If the probability that a person learning from a teacher
speaking grammar $G_{i}$ ends up with grammar $G_{j}$ is $M(j
\leftarrow i)$, the rate of change of the population speaking
$G_{j}$ can be written as 
\begin{equation}
\dot{X}(j,t)=\sum_{i=1}^n M(j \leftarrow i) W(\{X(i)\}) X(i,t)-\left(\sum_{i=1}^n W(\{X(i)\}) X(i,t) \right) X(j,t)
\label{conc}
\end{equation}
The interpretation of the terms in the above equation is similar to
(\ref{exdis}) or (\ref{exc}). However an important difference is that the
fitness $W(\{X(i)\})$ of the grammar $G_i$ now depends on the
frequency of the other grammars as well. Such a selection-mutation
equation with concentration-dependent fitness is known as 
replicator-mutator equation \cite{Nowak:2006}.    

Assuming that the error to any grammar is equally likely, it follows
that $M(j \leftarrow i)= q \delta_{i,j}+ \left[(1-q)/(n-1) \right]
(1-\delta_{i,j})$ where $q=1-\mu$ is the learning accuracy. A detailed analysis of the above equation is possible for the fitness choice  \cite{Komarova:2001}:
\begin{eqnarray}
 w(i,j)&=&w(j,i)=w \text{ for } i\neq j\\
 w(i,i)&=&1
\end{eqnarray}
\begin{figure} 
\begin{center}
\includegraphics[width=0.7\linewidth,angle=0]{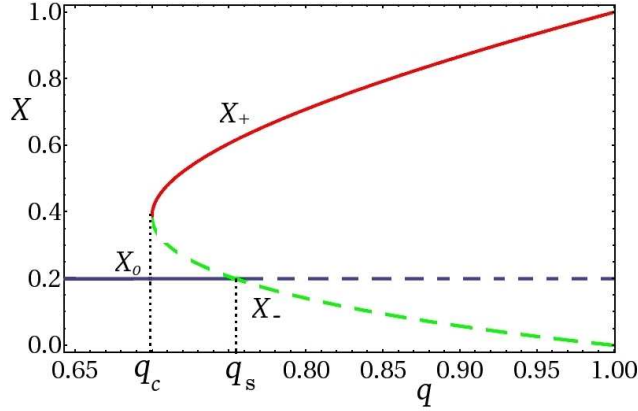} 
\caption{Multiple solutions of population frequency $X$ for $w=0.2$ and $n=5$
  (see Sec.\ref{concdep}). Unstable solutions are indicated by broken lines and stable ones by solid lines. } 
\label{grammar}
\end{center}
\end{figure}
The stable fixed points for the system of equations given by (\ref{conc}) can be found
by setting the left hand side to be zero and choosing all grammars
except one, say  $X(1)=X$, to be equally used so that,
$X(i)=(1-X)/(n-1),~i \neq 1$. This reduces the equation for $X(1)$ to  
\begin{equation}
 X^{3}-X^{2}q+\dfrac{(1-X)^{2}}{n-1}\left(X-\dfrac{1-q}{n-1} \right)+\dfrac{(1-q)
 w (nX-1)}{(1-w)(n-1)}=0  
\end{equation}
The above cubic equation for $X$ has three solutions namely
$X_{0},X_{+}$ and $X_{-}$ as shown in
Fig.~\ref{grammar}. The 
solution $X_0$ corresponds to the case in which all the
grammars are equally used and exists for all
$0 \leq q \leq 1$. The other two solutions  $X_\pm$ appear 
beyond a critical learning accuracy $q_c$ and correspond 
to the most used ($X_+$) and the least used ($X_-$) grammars.  Using a
linear stability analysis it can be shown that 
%$X_{-}$ is unstable for all $q \geq q_c$. If the 
%solutions $X_0$ and $X_-$ intersect at $q_s$,    the fraction $X_0$
%becomes unstable  for  $q>q_s$ and only the $X_{+}$ solution remains
%stable \cite{Komarova:2001,Nowak:2006}. Thus 
the stability of these 
solutions falls in three regimes depending on the learning accuracy
$q$: when $q < q_{c}$, the fraction $X_0$ is the only solution and
is stable, whereas
in the range $q_{c} \leq q < q_{s}$ all the three solutions exist but
$X_{-}$ is unstable and finally when $q \geq q_{s}$, the fraction $X_{0}$
also loses stability and $X_{+}$ is the only stable solution. 

Concentration based fitness is confined not just to languages but is
also seen in other systems such as host-parasite \cite{Brumer:2004,
  Sardanyes:2007} and immune system-pathogen interactions
\cite{Izmailian:2007, Kamp:2002}. In these systems, the
evolution is not based on the concentration of the same species
populations but on the concentration of other species. Thus their
evolution equations are coupled and this is dealt with in the next section.

%--------------------------------------------------------------------------
\subsection{Coupled quasispecies models}
\label{coup}

A class of models in which the growth of  a population  depends on another population constitute an example of a set of 
  nonlinear evolution equations. Below
  we discuss two such models in some detail.  

\begin{figure} 
\begin{center}
\includegraphics[width=0.9\linewidth,angle=0]{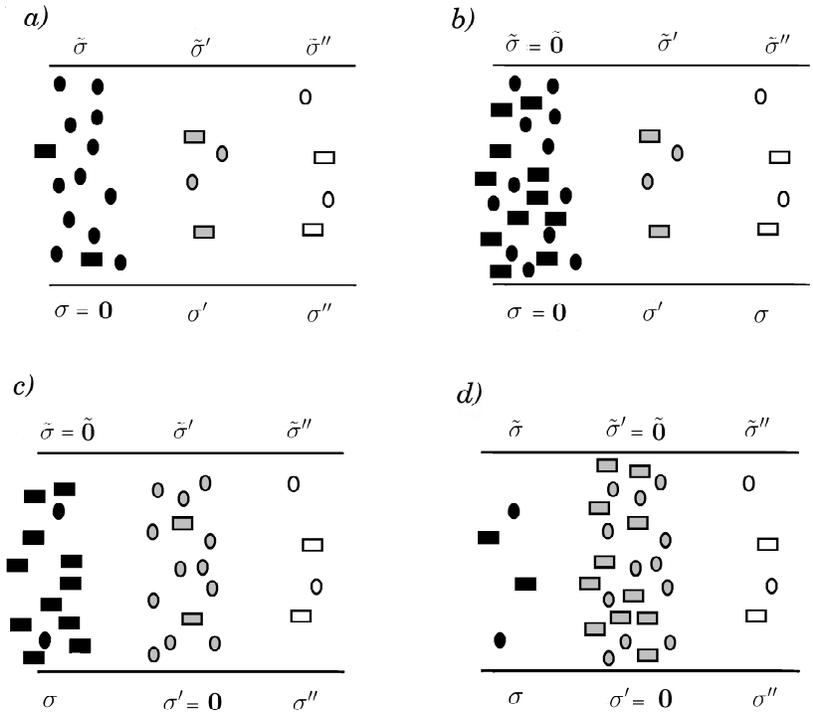} 
\caption{Dynamics of the coevolution of the viral
  (ovals) and the immune receptor sequences
  (rectangles): a) The viral quasispecies is initially formed 
  around a master sequence $\textbf{0}$ surrounded by its mutant
  sequences. b) The receptor sequence $\tilde {\bf 0}$ corresponding
  to $\textbf{0}$ proliferates and forms receptor quasispecies  around
  $\tilde {\bf 0}$. c)  
  To escape the attack of immune system, the viral master sequence 
  randomly shifts to one of its one mutant neighbours. d) In response, 
  the
  master sequence of the immune receptor also migrates to the location
  of the new viral master sequence.}
\label{coupled}
\end{center}
\end{figure}

{\it Coevolution of quasispecies:}
When an organism is infected by a virus, the immune receptors of the
host cell counterattack the virus. There is a one-to-one mapping
between the virus and the immune receptors so that a viral sequence $\sigma$ is
attacked only by its corresponding receptor sequence ${\tilde
  \sigma}$, $\sigma'$ only by ${\tilde
  \sigma}'$ and so on. 
In order
to escape the 
immune system, the virus adapts and in response, the immune system 
adapts to counter the new viral strain (see Fig.~\ref{coupled}) and this cycle repeats over a time period $\tau$. Thus the viral
species and the immune receptors are involved in a dynamic
evolutionary race but may coexist under certain conditions as
explained below. 

Assuming that both the receptor and viral sequences have the same length $L$, the evolution equations for the frequency $X({\tilde \sigma},t)$ of immune
receptor sequence ${\tilde \sigma}$ and $x(\sigma,t)$ of the
corresponding viral sequence ${\sigma}$ can be written as \cite{Kamp:2002}:  
\bea
 \dot{X}({\tilde \sigma},t) &=& \sum_{{{\tilde \sigma}'}}M_\nu({\tilde \sigma} \leftarrow
 {\tilde \sigma}') W (x(\sigma',t)) X({\tilde \sigma}',t)-{\tilde D} X({\tilde \sigma},t) \\
\dot{x}(\sigma,t) &=& \sum_{ ^{\sigma'}} M_\mu(\sigma \leftarrow
 \sigma') w (\sigma',t)x(\sigma',t) -D(X({\tilde \sigma},t)) x(\sigma,t) 
\eea
where the subscripts in the sequence mutation probability $M$ (see (\ref{mutprob1})) denote
the mutation probability per locus and the death term of the immune receptor ${\tilde D}=\sum_{
   {\tilde \sigma}'}W (x(\sigma',t)) X({\tilde \sigma}',t)$. As the immune receptor population moves in response to
the viral population, the fitness $W (x(\sigma,t))$ of the receptor $\tilde \sigma$ depends on the concentration of the corresponding viral sequence $\sigma$. In the above equations, the death terms are different for the two
 populations as the number of immune receptors is conserved while the
 virus number is not. For simplicity, one can choose the death rate of the virus as 
\begin{eqnarray}
D(X({\tilde \sigma},t)) =
\begin{cases}
\delta \text{ ,~if } {\tilde \sigma}= \text{immune receptor master sequence}\\
	0 \text{ ,~otherwise }
\end{cases}
\end{eqnarray}
A time-dependent sharp peak fitness landscape is assumed 
for both immune receptor and virus as their master sequences move
through the sequence space. Since the viral fitness $w$ is independent
of $X(\sigma)$, we can write 
\begin{eqnarray}
w(\sigma,t) =
\begin{cases}
w_0 \text{ ,~if } \sigma= \text{viral master sequence at time $t$}\\
	1 \text{ ,~otherwise }
\end{cases}
\end{eqnarray}
where $w_0 > 1$. 
Similarly $W (x(\sigma,t))=W_0>1$ if
$\sigma$ is the viral master sequence and unity otherwise.
%given by 
%\begin{eqnarray}
%W (x(\sigma,t))&=&\alpha\text{ ,~if } \sigma= \text{viral master sequence}\\
%&=&\Delta<< \alpha\text{ ,~otherwise }
%\end{eqnarray}}

In periodically changing fitness landscapes such as being
  considered here, there is no steady state as the population keeps
  migrating with the fitness landscape. However one can still define
  an error threshold in the large time limit analogous to that on
  static fitness landscapes as the maximum mutation rate above which
  the population gets uniformly distributed over the sequence space. A
  possible way to determine the critical mutation rate is to consider
  the behavior of relative frequency $\kappa$ of the new master
  sequence  to the frequency of a sequence far away from the current
  master sequence at the time period $\tau$ of the fitness landscape
  \cite{Nilsson:2000}. At large times, it is a good approximation to
  assume that the far-off sequences have reached a quasi-equilibrium
  and therefore their unnormalised frequency grows exponentially fast
  with the growth rate given by the respective fitness. However such
  an equilibrium is not reached for the populations in the vicinity of
  the (migrating) master sequence and the growth at such sequences
  depends on the mutational contribution from the current master sequence. If the mutation probability or the time period is too small, the population cannot build up at the new master sequence and the relative frequency $\kappa < 1$. On the other hand,  the new master sequence grows for $\kappa > 1$. Thus $\kappa=1$ marks the transition point between the extinction and survival phases of the quasispecies on periodically changing fitness landscapes. 

%An approximate analysis of the above equations shows thatIf the viralquasispecies adapts to the new fitness peak over a time scale $t_0$and the migrating immune system over $T_0$, the cycle described in
%Fig.~\ref{coupled} repeats itself over $T=t_0+T_0$ time scale. {\it
  %how times determined? meaning of kappa} \textbf{As this system does
  %not have a steady state, the growth of population is determined by
  %the fraction $\kappa$ which measures the growth of the 
 % master sequence relative to a background sequence located far away from the
  %master sequence. If $\kappa$ is
%above unity, the master sequence evolves faster than the rest and the population survives else it goes extinct.}%at the new master sequence relative to the background sequences which is above unity for surviving population and below one for extinct population.%
Following the arguments sketched above, the fraction $\kappa_\mu$  for
the virus can be found and is given by \cite{Kamp:2002}
\be
\kappa_\mu=\frac{\left(e^{((1-\mu)^L w_0-1) \tau}-e^{((1-\mu)^L-1) \tau} \right) \mu w_0}{(w_0-1) (1-\mu)}
\label{kappa}
\ee 
The relative frequency of the immune receptors $\kappa_\nu$ is obtained on replacing $\mu$ by $\nu$ and $w_0$ by
$W_0$ in the above expression. 
%Using the approximate expressions for the time $t_0 (\mu,\delta)$ and
%$T_0(\nu)$ in (\ref{kappa}) and s
Setting $\kappa_\mu$ and $\kappa_\nu$ equal to one gives a phase diagram in $\mu-\nu $ plane which shows that 
while both the
populations exhibit the classical error catastrophe at high mutation
rates (as discussed in Sec.~\ref{hap}), the viral
population has an additional transition point when its mutation rate
is too low to escape the immune response and in between these values the two populations coexist
  \cite{Nilsson:2000,Kamp:2002}. The predicted mutation rates
of the B-cells that produce the immune receptors and the receptor
lengths that maximise both regimes of viral error catastrophe for
optimal immune response are seen to match the experimental
observations \cite{Kamp:2002}.

\begin{figure} 
\begin{center}
\psfrag{P=1}{$x=1$}
\psfrag{P,Q}{$x, X$}
\psfrag{Q=1}{$X=1$}
\includegraphics[width=0.5\linewidth,angle=0]{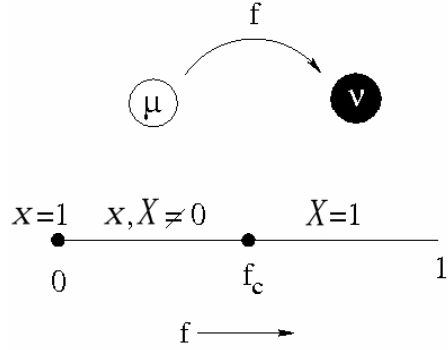} 
\end{center}
\caption{Schematic phase diagram of the quasispecies model with
  nonmutator and mutator populations. The pure nonmutator phase occurs
  when $f=0$, 
  pure mutator phase for $f \geq f_c$ and the system is in the  mixed
  phase for $0<f<f_c$ (see Sec.~\ref{coup})}
\label{fphase}
\end{figure}

{\it Evolution of a mixed population:} As discussed in Sec.~\ref{hap},
there exists an error 
threshold above which the mutational load is too high to be
compensated by selection. 
 For this reason, and because most
mutations are known to have deleterious effect
\cite{Sturtevant:1937,Drake:1998}, the 
spontaneous mutation rate is expected to be small
\cite{Kimura:1967}.  However small subpopulations of strains with
high mutation rates have been observed in natural isolates
\cite{LeClerc:1996} and in
experiments \cite{Sniegowski:1997,Boe:2000}.

Consider such a mixed population with nonmutator and mutator strains with
mutation probability $\mu$ and  $\nu=\lambda \mu, \lambda > 1$
respectively. Due to the damage in error repair systems, the mutation
rate of normal strains can rise and hence a nonmutator can convert
to a mutator  with probability
$f$ . 
Then the average fraction $x(\sigma,t)$ and $X(\sigma,t)$ of the nonmutator
and the mutator respectively at generation $t$ evolves 
according to the 
following coupled nonlinear difference equations \cite{Nagar:2009}:
\bea
x(\sigma,t+1) &=& \frac{(1-f) \sum_{\sigma'} M_\mu (\sigma \leftarrow
  \sigma') W(\sigma') 
  x(\sigma',t)}{{\cal W}(t)} {}  \label{Pk}\\ 
X(\sigma,t+1) &=& \frac{\sum_{\sigma'} M_\nu (\sigma \leftarrow
  \sigma') W(\sigma') 
  X(\sigma',t)}{{\cal W}(t)} \nonumber \\
&+&\frac{f \sum_{\sigma'} M_\mu(\sigma
  \leftarrow \sigma') W(\sigma') 
  x(\sigma',t)}{{\cal W}(t)} ~.\label{Qk}
\eea
where the average fitness ${\cal W}(t)
=\sum_{\sigma} W(\sigma) \left[x(\sigma,t)+
  X(\sigma,t) \right]$ and the subscripts in the mutation matrix refer
to the mutation probability per locus per generation. For the
  reasons mentioned above, the mutators are selected against and their
  number is expected to be low. But with increasing $f$, mutators are continually generated thus increasing their frequency and at sufficiently high $f$, the mutator
  population can reach unity. Thus a phase transition can occur at a 
critical probability $f_c$ between the mixed phase with both
nonmutator and mutator 
population and a pure mutator phase (see Fig.~\ref{fphase}). In the
steady state, such a
phase transition has been shown to occur on single peak fitness
landscapes \cite{Tannenbaum:2003} and multiplicative fitness landscapes
\cite{Nagar:2009}.
  
To see this transition for fitness choice (\ref{mult1}), we first observe that $x(\sigma)=0$ is a solution
of Eq.~(\ref{Pk}) and thus corresponds to a phase in which the entire 
population consists of mutators and the total mutator fraction
$X=\sum_\sigma X(\sigma)=1$. As  
(\ref{Qk}) reduces to (\ref{exdis}) in this phase, using the exact solution 
(\ref{prod}), the average fitness ${\cal W}_>$ in the $f > f_c$ phase
can be found.  
If, on the other hand,  the total 
nonmutator fraction  $x=\sum_\sigma x(\sigma)$ is nonzero,
on summing over all the sequences on both sides of  
Eq.~(\ref{Pk}), we find that the average fitness
${\cal W}_<$ in the mixed phase corresponding to $f < f_c$ does not
depend on the mutator fraction and can be written as 
\be
{\cal W}_<=\frac{(1-f) \sum_\sigma W(\sigma) x(\sigma)}{\sum_\sigma
  x(\sigma)}~,~x \neq 0 
\label{fitreln}
\ee
thus leading to an uncoupled nonlinear equation for
$x(\sigma)$. On eliminating ${\cal W}$ from Eq.~(\ref{Pk}) using the above
equation, we see that $x(\sigma)/\sum_{\sigma'}
x(\sigma')$ obeys the quasispecies 
equation (\ref{exdis}) and one can find the average fitness ${\cal
  W}_<$ as well.  Equating the fitnesses ${\cal W}_<$ and ${\cal W}_>$
at the critical point,  the 
phase boundary in the $f-\lambda$ plane is obtained,   
\be
(1-f_c)^{1/L}
=\frac{(2-s) (1-\nu_c)+\sqrt{4 \nu_c^2 (1-s)+s^2
    (1-\nu_c)^2}}{(2-s)(1-\mu)+\sqrt{4\mu^2(1-s)+s^2(1-\mu)^2}} 
\label{phasedia}
\ee 
Using the above analysis, it is also possible to calculate the average
mutator fraction as a function of $f$ and $\lambda$. The results are
seen to be in good agreement with the experiments on {\it E. coli}
\cite{Nagar:2009}.  

%=============================================================================
%SEXUAL 
%=============================================================================
\section{Sexually reproducing populations}
\label{sex}

In this section, we mainly consider a recombining haploid population
with sequence length two. As discussed in Sec.~\ref{basic}, due to
recombination, the sequences $\{0,0\}$ and 
$\{1,1\}$ can give rise to offspring sequences $\{ 0,1\}$ or 
  $\{1,0\}$. Similarly the recombination between  $\{0,1\}$ and 
  $\{1,0\}$ can result in  $\{0,0\}$ and $\{1,1\}$. In the following, for
brevity we denote the population at the sequences $\{0,0\},
\{0,1\},\{1,0\}$ and $\{1,1\}$ by $X_0, X_1, X_2$ and $X_3$ 
 and their respective fitness by $W_0, W_1, W_2$ and $W_3$. 
If such a population  undergoes recombination alone, the
  frequency $X_i(t)$ evolves according to the following equation:
\be
X_i(t+1)= \sum_{j,k=0}^{3} R(i \leftarrow j,k) X_j (t) X_k(t)
\label{rec1}
\ee
where $R(i \leftarrow j,k)$ is the probability that a sequence $i$ is obtained by recombining sequences $j$ and $k$. The recombination process between suitable sequences is assumed to occur with probability $r$ and does not occur with $1-r$.  For example, for the offspring sequence $0$, we have 
\bea
&&R(0 \leftarrow 0,0)=1 ~,~ R(0 \leftarrow 0,1)=R(0 \leftarrow 0,2)=\frac{r}{2}+\frac{1-r}{2}=\frac{1}{2}\\
&&R( 0 \leftarrow 0,3)=\frac{1-r}{2} ~,~R( 0 \leftarrow 1,2)=\frac{r}{2}
\eea
and the rest of the probabilities are zero. On writing the recombination probabilities in a similar manner for other sequences  and using the normalisation $\sum_{i=0}^3 X_i=1$, we find that  
the population fractions evolve according to \cite{Ewens:1979}
\bea
X_0(t+1) &= &X_0(t)+r (X_1(t) X_2(t)-X_0(t) X_3(t)) \label{basicrec1} \\
X_1(t+1) &= &X_1(t)+r (X_0(t) X_3(t)-X_1(t) X_2(t)) \\
X_2(t+1) &= &X_2(t)+r  (X_0(t) X_3(t)-X_1(t) X_2(t)) \\
X_3(t+1) &= &X_3(t)+r (X_1(t) X_2(t)-X_0(t) X_3(t))
\label{basicrec4}
\eea
%where $r$ is the recombination probability. 
Thus the population fractions obey a set of nonlinear equations when
recombination is present and it is not known if these equations can be
linearised. 

The bilinear frequency combination $X_1(t)
  X_2(t)-X_0(t) X_3(t)$ is called {\it linkage disequilibrium} $D(t)$
  at time $t$ and is a measure of the correlation between the
  frequency at the two loci. Using
  (\ref{basicrec1})-(\ref{basicrec4}) we
  have $D(t+1)= (1-r) D(t)$ so that the linkage disequilibrium  
  vanishes in the steady state {\it i.e.} $X_1 X_2=X_0 X_3$ and as
  a consequence, the frequency of the
  sequence $\{\sigma_1, \sigma_2\}$ equals the product of frequency of
  sequences $\{ \sigma_1 \}$ and $\{ \sigma_2 \}$. For example, the
  frequency of zero sequence at first locus equals
  $X_0+X_1$ and that at the second locus is $X_0+X_2$. Using $D=0$, it follows that
  the product $(X_0+X_1) (X_0+X_2)=X_0$, the frequency of the sequence
  $\{0,0\}$. Although the linkage
  disequilibrium is zero when only recombination is present, it is
  usually nonzero when selection and/or mutation
are also included.

We now discuss the situation when selection, mutation
and recombination are present. We will consider the fitness scheme in
which two fitness 
peaks are separated by a fitness valley and assume that $W_3 > W_0=1 >
W_1=W_2 $. In a population initially localised at $\{0,0\}$, a
mutation in $\{0,0\}$ to $\{0,1\}$ or $\{1,0\}$ is  
deleterious but the fitness loss can be compensated by acquiring another
mutation resulting in the sequence $\{1,1\}$.   In the absence of
recombination and for small mutation rates, the population will
eventually localise around the fittest $\{1,1\}$ sequence (see
Sec~\ref{hap}). However due to nonlinear evolution equations,  
multiple steady states may result \cite{Crow:1965,Higgs:1998}.  
As discussed below, there exists a critical recombination rate $r_c$
below which the population can cross the intervening valley and reach
the fittest peak at $\{1,1\}$. But above $r_c$, the population can
remain trapped at the initial sequence with low fitness and thus the
sexual reproduction can affect the adaptation process adversely. We
now describe the population behavior for two schemes of mutation rates. 
%but the behavior above $r_c$ depends onthe choice of mutation rates \textbf{as explained below}. 

%--------------------------------------------------------------------------
{\it Multiple equilibria in steady state:}
If the mutation matrix is symmetric and given by (\ref{mutprob1}), 
the evolution equations can be
written as \cite{Park:2010}
\bea
X_0(t+1)&=& X_0'(t)- r (1-2 \mu)^2 \frac{D(t)}{{\cal W}^2(t)} 
\label{multrec1}\\
X_1(t+1) &=& X_1'(t)+ r (1-2 \mu)^2 \frac{D(t)}{{\cal W}^2(t)} \\
X_2(t+1) &=& X_2'(t)+r (1-2 \mu)^2 \frac{D(t)}{{\cal W}^2(t)} \\
X_3(t+1) &=& X_3'(t)- r (1-2 \mu)^2 \frac{D(t)}{{\cal W}^2(t)} 
\label{multrec4}
\eea
where ${\cal W}(t)=\sum_{k=0}^3 W_k X_k(t)$ is  the average
fitness of the population, the linkage disequilibrium $D(t)=W_0 W_3 X_0(t) X_3(t) - W_1 W_2 X_1(t)
X_2(t)$ and the primed fractions are given by the left hand side of
(\ref{exdis}):
\be
X_i'(t)=\frac{\sum_{j=0}^3 
M(i\leftarrow j) W_j X_j(t)}
{{\cal W}(t)}
\ee
 To arrive at the set of equations (\ref{multrec1})- (\ref{multrec4}), it has
been assumed that recombination occurs after selection and
mutation. Thus in the set of equations (\ref{basicrec1})-(\ref{basicrec4}), the frequency on the right hand side refers to $X_i'(t)$ upon using which (\ref{multrec1})- (\ref{multrec4}) are obtained.

In the steady state, for the fitness landscape described
above, the fractions $X_i$'s can be expressed in terms of the fitness
$W_i$'s and the average fitness ${\cal W}$. On using the resulting
expressions for $X_i$'s in the equation for ${\cal W}$, a quartic
equation for ${\cal W}$ is 
obtained.  
An analysis \cite{Park:2010} of this equation shows that for $r <
r_c$, the fittest 
sequence is always populated while for $r > r_c$, 
there are two stable solutions: either the population stays at the
initial sequence $\{0,0 \}$ or moves to the fittest sequence
$\{1,1\}$.  

%--------------------------------------------------------------------------
{\it Time to fixation:} If the mutations are unidirectional with the
probability to mutate from $0$ to $1$ being $\mu$ and zero for the
back mutation, the whole population occupies the fittest sequence and
the sequence $\{1,1\}$ is said to be fixed. In
such a case, it is interesting to study the dynamics of the
population and more specifically, one can find the time $T$ to fixation.  
 
For the one-way mutation scheme in which first selection takes place followed by recombination and finally mutation, the time evolution occurs according 
to the following nonlinear equations \cite{Jain:2010a}: 
\bea
X_0(t+1)&=& \frac{ (1-\mu)^2 W_0 X_0(t)- r (1-\mu)^2 D(t)}{{\cal W}(t)} \\
X_1(t+1) &=& \frac{ \mu (1-\mu) W_0 X_0(t)+  (1-\mu) W_1 X_1(t)+r (1-\mu)^2 D(t)}{{\cal W}(t)} \\
X_2(t+1) &=& \frac{ \mu (1-\mu) W_0 X_0(t)+  (1-\mu) W_2 X_2(t)+r (1-\mu)^2  D(t)}{{\cal W}(t)} \\
X_3(t+1) &=& \frac{\mu^2 W_0 X_0(t) + \mu (W_1 X_1(t)+ W_2 X_2(t)) +
  W_3 X_3(t)}{{\cal W}(t)} \nonumber \\
&-&\frac{r
  (1-\mu)^2  D (t)}{{\cal W}(t)}
\eea
Here $D(t)=(W_0 W_3 X_0(t) X_3(t)-W_1 W_2 X_1(t) X_2(t))/{\cal W}(t)$ is the linkage disequilibrium
at time $t$ and ${\cal W}(t)=\sum_{k=0}^3 W_k X_k(t)$ is  the average
fitness of the population. The above equations can be written
  down in a manner analogous to the above cases. Since selection
  occurs before recombination, on replacing $X_i(t)$ by $W_i
  X_i(t)/{\cal W}(t)$ on the RHS of
  (\ref{basicrec1})-(\ref{basicrec4}), the evolution equations with
  selection and recombination are obtained. Finally the unidirectional
  mutation scheme is implemented.

The equations for the corresponding unnormalised populations $Z_k$'s
defined by (\ref{Eigendisc}) can also be written. But due to the
recombination term, the equations for $Z_k$'s also remain
nonlinear. An approximate method to handle these dynamical nonlinear
equations can be developed by noting that at any instant, for small
mutation rates, only one of the four populations dominate. Then the
dynamics of  population $Z_k$'s can be divided in following three dynamical
phases \cite{Jain:2010a} : 
(i) $Z_0 \gg Z_1, Z_3$ (phase I) (ii) $Z_1 \gg Z_0, Z_3$ (phase
II) and (iii) $Z_3 \gg Z_0, Z_1$ (phase III). Thus one can expand the 
equations for unnormalised populations in powers of $Z_1/Z_0, Z_3/Z_0$
in phase I, $Z_0/Z_1, Z_3/Z_1$ in phase II and similarly, $Z_0/Z_3,
Z_1/Z_3$ in phase III.  The time at which a
phase ends is obtained by matching the solutions of the relevant
populations in the two phases. The fixation time is then obtained by
summing over these phase times.   

\begin{figure}
\includegraphics[width=0.6 \linewidth,angle=270]{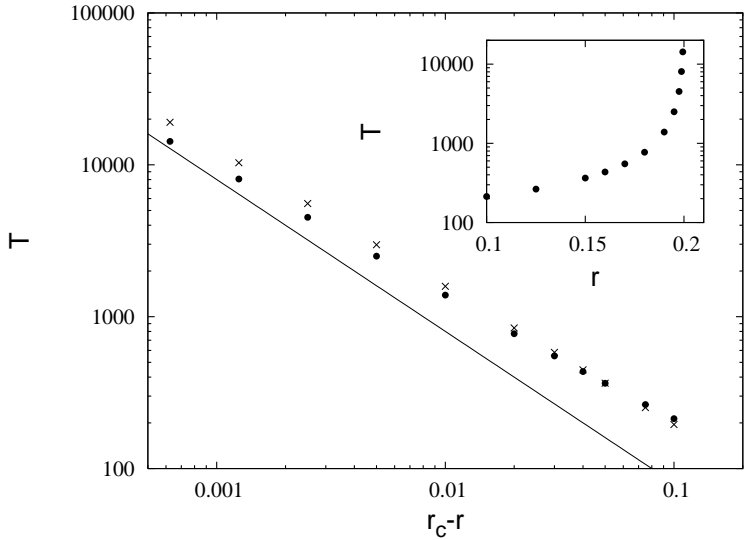}
\caption{Fixation time as a function of $r$
  obtained using exact 
  iteration ($\bullet$) and analytical result ($\times$) given by 
  (\ref{tau1comp2}). The
  solid line has a slope equal to $-1$.}
\label{comptime}
\end{figure}

As mentioned above, there exists a critical recombination fraction $r_c$ 
beyond which a 
population initially located at $\{0,0\}$ cannot cross the intermediate
fitness valley and reach the double mutant
fitness peak \cite{Crow:1965,Eshel:1970}.  The
inset of Fig.~\ref{comptime} shows that the 
fixation time diverges as $r$ approaches critical recombination
probability $r_c=(w_4-1)/w_4$.
A simple calculation  using the method described above but ignoring
the nonlinearities shows that the 
fixation time diverges as $1/(r_c-r)$.  However a more careful
analysis that takes the nonlinear terms into account shows that the
fixation time is well approximated by \cite{Jain:2010a} 
\be
T \approx \frac{1}{(r_c-r) W_3} \left[ \ln \ln \left(\frac{W_3 (1-W_1)
    (r_c-r)^2}{(1-W_1+r_c W_3) W_1^2 K^{2}} \right)-\ln
  \left(\frac{2 r_c^2 W_1^2K}{(1-W_1) (r_c-r)^2}\right) \right]
\label{tau1comp2}
\ee
where the constant $K \sim (r_c-r)^{-1}$. Thus the fixation time
decays slower than $1/(r_c-r)$ due to the logarithmic corrections (see
Fig.~\ref{comptime}).   

The population frequencies and fixation time can be analysed for other
fitness schemes as well and a discussion can be found in
\cite{Eshel:1970,Jain:2010a}. 
Although we have discussed the haploid
case, the diploid problem has also been studied  \cite{Burger:2000}. 
For studies on models that consider more than two loci, the reader may
refer to \cite{Otto:1997b,Jacobi:2006}.  

%=============================================================================
%SUMMARY
%=============================================================================
\section{Summary}
\label{sum}

In this review, we have presented a brief (and incomplete) overview of
evolutionary 
processes and models in deterministically evolving 
populations. As we have discussed, these systems are inherently
nonlinear and difficult to analyse analytically. The nonlinearity
of these systems that makes them so difficult to handle, is also
responsible for the complex behaviour of their solutions. The
existence of multiple
steady states and dynamic phase transitions are some of the interesting 
features displayed by these models. 

While these theoretical models of evolutionary biology have garnered 
interest amongst physicists and mathematicians, they have also been 
successful in predicting biological properties and explaining the
experimental results quantitatively.  It is hoped that the work 
integrated from various disciplines will take us closer to an 
understanding of the complex and 
continuous process of the evolution of life. 

%=============================================================================
%BIBLIOGRAPHY
%=============================================================================

%=============================================================================
\end{document}